\documentclass[aps,pre,onecolumn,superscriptaddress,showpacs]{revtex4-1}
\usepackage{graphicx}
\usepackage{dcolumn}
\usepackage{bm}

\usepackage{natbib}
\usepackage[
text={7in,9.5in},centering,
]{geometry}

\usepackage{float}
\usepackage{epsfig}
\usepackage{amsmath}
\usepackage{amssymb}
\usepackage{textcomp}
\usepackage{layouts}

\begin{document}

\title{A Way of Separating Dynamics and Gauge Transformations in Yang-Mills Theory}

\author{Natalia Gorobey}
\affiliation{Peter the Great Saint Petersburg Polytechnic University, Polytekhnicheskaya
29, 195251, St. Petersburg, Russia}

\author{Alexander Lukyanenko}\email{alex.lukyan@mail.ru}
\affiliation{Peter the Great Saint Petersburg Polytechnic University, Polytekhnicheskaya
29, 195251, St. Petersburg, Russia}

\author{A. V. Goltsev}
\affiliation{Ioffe Physical- Technical Institute, Polytekhnicheskaya
26, 195251, St. Petersburg, Russia}

\begin{abstract}
A modification of the gauge theory is proposed, in which the set of generalized coordinates is supplemented with symmetry transformation parameters, and a condition is additionally imposed on the latter that ensures the classical character of their dynamics in quantum theory. As a result, additional dynamic variables and transverse physical degrees of freedom in the Hamiltonian are separated. The classical theory of the Yang-Mills field is considered.

\end{abstract}


\maketitle

\section{Introduction}
A simple and natural understanding of the dynamic structure of the gauge theory is that the initial set of dynamic variables in it is redundant and some of them should be eliminated by solving the equations of constraints and additional gauge conditions \cite{1}. However, in the subsequent development, this reduction procedure was replaced by the expansion of the phase space of the theory with addition of Lagrange multipliers with the corresponding canonical momenta and ghosts, as well as the expanded BRST - symmetry \cite{2,3,4,5,6}. Another option for expanding the phase space of the gauge theory was proposed in \cite{7}, where the parameters of finite symmetry transformations are added to the original dynamic variables. These finite shifts in the group space are constructed in the form of integrals of infinitesimal shifts generated by constraints. However, such an extension alone does not solve the problem of separating physical degrees of freedom and pure calibrations and a dynamic interpretation of the theory. It should be supplemented with a structure that allows connecting finite shifts in the group space with observations. In the case of the dynamics of relativistic particles with reparametrization invariance of their world lines, the intrinsic parameter of the symmetry group plays the intrinsic time of each particle. This invariant parametrization also arises naturally in the BRST-invariant representation of the propagator of covariant quantum theory for a relativistic particle \cite{8} and reproduces the Fock \cite{9} and Schwinger \cite{10} formalism based on the introduction of proper time.
In \cite{11}, the introduction of this parameter was proposed to be supplemented with the condition of its classical dynamics with the corresponding modification of the initial action. This addition allows us to connect the proper time with observations and get a dynamic interpretation of covariant quantum theory. It can be assumed that such two-stage modification of the singular theory (adding finite symmetry transformations to dynamic variables and an additional condition for their classicality) will be an effective way to separate physical degrees of freedom and pure gauges in the general case. In this paper, this will be shown as an example of a free Yang-Mills field. The result will be a separation of the dynamics of the physical transverse components of the Yang-Mills field and the `motion' of the longitudinal components (pure calibrations) in the group space.

\section{Modification of the action of Yang-Mills}
The proposed modification of the gauge theory action is divided into two stages. We proceed from the canonical form of action
\begin{equation}
 I=\int{dt}[p_i\dot{q}_i-\lambda_a\phi_a(p,q)-h(p,q)],
 \label{1}
\end{equation}
where the constraints obey the commutation relations
\begin{equation}
\left\{{\phi_a,\phi_b}\right\}=C_{abd}\phi_d, \left\{{\phi_a,h}\right\}=h_a\phi_a.
 \label{2}
\end{equation}
(we consider the case $h_a=0$), and variations of the Lagrangian multipliers ensuring the invariance of the action (\ref{1}) with respect to the infinitesimal symmetry transformations
\begin{equation}
\delta{q_i}=\epsilon_a\left\{q_i,\phi_a\right\}, \delta{p_i}=\epsilon_a\left\{p_i,\phi_a\right\}
  \label{3}
\end{equation}
have the form
\begin{equation}
 \delta{\lambda_a}=\dot{\epsilon}_a-C_{abd}\lambda_b\epsilon_d.
  \label{4}
\end{equation}
At the first stage, according to \cite{7}, we replace the Lagrangian multipliers with explicit functions of the parameters defining the finite symmetry transformation ($\delta{s}_a=\epsilon_a$):
\begin{equation}
 \lambda_a=\dot{s}_b\Lambda_{ba}(s).
  \label{5}
\end{equation}
which are integrals of functional-differential equations (\ref{4}).
At the second stage, according to \cite{11}, we add to the action a variation generated by the infinitesimal shift of new dynamic variables. We call this as a condition of classicality, since it allows one to remove integration over new dynamic variables in the functional-integral representation of the propagator of covariant quantum theory.
We carry out these constructions as an example of a free Yang-Mills field $A_{\mu a}$, where $\mu=0,1,2,3$ is space-time index, and $a$ is an internal index of the gauge theory. $A_{0a}$ are the Lagrange multipliers here, so at the first stage, the original Yang-Mills Lagrangian function takes the form: in the first stage
\begin{equation}
L=\frac{1}{2}[(\dot{A}_{ia}-\nabla_i(\dot{s}_b\Lambda_{ba}))^2-B_{ia}^2]
 \label{6}
\end{equation}
where $B_{ia}$ is the Yang-Mills `magnetic field' tension, and the covariant derivative is determined by the relation \cite{1}:
\begin{equation}
\nabla_i F_a=\partial_i F_a-igT_{bad}A_{ib} F_d.
 \label{7}
\end{equation}
We will not need the explicit form of functions $\Lambda_{ab}$ here.
Now, following \cite{8}, we will still expand the set of new variables by adding infinitesimal shifts $\epsilon_a$ to them, and to Lagrangian function (\ref{6}) we add its variation generated by these infinitesimal shifts:
\begin{equation}
\tilde{L}=\frac{1}{2}[(\dot{A}_{ia}-\nabla_i(\dot{s}_b\Lambda_{ba}))^2-B_{ia}^2]+(\dot{A}_{ia}-\nabla_i(\dot{s}_b\Lambda_{ba}))(\nabla_i(\dot{\epsilon}_c\Lambda_{ca}+\dot{s}_c\frac{\partial\Lambda_{ca}}{\partial{s_d}}\epsilon_d),
  \label{8}
\end{equation}

\section{The canonical form of the modified Yang-Mills action}
We turn to the canonical form of the modified action (\ref{8}). We find the canonical momenta:
\begin{equation}
\pi_{ia}=\dot{A}_{ia}-\nabla_i(\dot{s}_b\Lambda_{ba})+\nabla_i(\dot{\epsilon}_c\Lambda_{ca}+\dot{s}_c\frac{\partial\Lambda_{ca}}{\partial{s_d}}\epsilon_d),
  \label{9}
\end{equation}
conjugated to $A_{ia}$,
\begin{equation}
p_{s_b}=-\Lambda_{ab}\nabla_i(\dot{A}_{ia}-\nabla_i(\dot{s}_c\Lambda_{ca}))-\nabla_i(\dot{A}_{ic}-\nabla_i(\dot{s}_q\Lambda_{qc}))\frac{\partial\Lambda_{bc}}{\partial{s_d}}\epsilon_d
-\Lambda_{ab}\Delta(\dot{\epsilon}_c\Lambda_{ca}+\dot{s}_c\frac{\partial\Lambda_{ca}}{\partial{s_d}}\epsilon_d),
 \label{10}
\end{equation}
conjugated to $s_a$ ($\Delta=\nabla_i\nabla_i$), and
\begin{equation}
 P_{\epsilon_b}=-\Lambda_{ab}\nabla_i(\dot{A}_{ia}-\nabla_i(\dot{s}_c\Lambda_{ca})),
  \label{11}
\end{equation}
conjugated to $\epsilon_b$.
From these momenta we immediately obtain the constraint equations,
\begin{equation}
p_{s_{b}} =-\Lambda_{ab} \nabla_{i} \pi_{ia} + P_{\epsilon_{a}} \Lambda_{ac}^{-1} \frac{\partial\Lambda_{cb}}{\partial{s_d}}\epsilon_d,
  \label{12}
\end{equation}
and generalized velocities in the following combination:
\begin{equation}
\dot{\epsilon}_c\Lambda_{ca}+\dot{s}_c\frac{\partial\Lambda_{ca}}{\partial{s_d}}\epsilon_d={\Delta}^{-1}(\nabla_i\pi_{ia}+\Lambda_{ab}^{-1}p_{\epsilon_b}) .
  \label{13}
\end{equation}
Now we find the Hamilton function of the modified theory:
\begin{equation}
 \tilde{h}=\frac{1}{2}[\pi_{ia}^2+B_{ia}^2]-\frac{1}{2}[\nabla_i\Delta^{-1}(\nabla_k\pi_{ka}+\Lambda_{ab}^{-1}p_{\epsilon_b})]^2.
  \label{14}
\end{equation}
where we used (\ref{13}).
Let's see what we got as a result. Obviously, the constraints (\ref{12}) commute with the Hamiltonian (\ref{14}). The Hamiltonian does not contain $\epsilon_a$. It means that canonical momenta $P_{\epsilon_a}$ (density of a color charge) are integrals of motion. We perform the orthogonal longitudinal-transverse splitting of the canonical momenta:
\begin{equation}
 \pi_{ia}=\nabla_i(\chi_{0a}^{L}+\chi_a^L)+\pi_{ia}^T,
  \label{15}
\end{equation}
with
\begin{equation}
\Lambda_{ab}\Delta\chi_{0b}^{L}=-P_{\epsilon_a}.
\label{16}
\end{equation}
As a result, the quadratic form of the momenta in the Hamiltonian contains only the transverse components:
\begin{equation}
 \tilde{h}=\frac{1}{2}[{\pi_{ia}^T}^2+B_{ia}^2].
  \label{17}
\end{equation}
Thus, the longitudinal components of the Yang-Fills are completely excluded from the dynamics in time. For them, only the ``dynamics'' in the group space, described by the constraints (\ref{12}), remains. Here, the evolution parameters $s_a$ are supplemented by dynamic variables $P_{\epsilon_a}$
which can be eliminated by choosing the origin of the longitudinal component of the momentum according to (16). In a gauge theory with the constraints linear in canonical momenta, these quantities do not have a dynamic meaning.

\section{Conclusions}
Thus, in the Yang-Mills theory, and generally in the theory with the constraints linear in canonical momenta, the introduction of the classical parameters of symmetry transformations as additional dynamic variables allows us to separate the physical transverse and gauge longitudinal degrees of freedom. At the same time, classical external sources, which are generators of classical symmetry transformations, are also added as dynamic variables. These sources themselves can be set equal to zero, as long as the separation of the physical degrees of freedom is done. In theories with quadratic on the canonical momenta constraints, such as the theory of gravity, in which there is a time problem, the modification proposed here introduces the concept of proper time, which also has its own classical source - energy.
In contrast to the case considered here, this energy can have a dynamic meaning. This issue will be considered separately.

\section{acknowledgements}
The authors thank V.A. Franke for useful discussions.

\end{document}